\documentclass{ws-mplb}

\begin{document}

\markboth{Sei-ichiro Suga and Kensuke Inaba}{COLOR SUPERFLUID AND TRIONIC STATES AT FINITE TEMPERATURES}

%
\catchline{}{}{}{}{}
%

\title{COLOR SUPERFLUID AND TRIONIC STATE OF ATTRACTIVE THREE-COMPONENT LATTICE FERMIONIC ATOMS AT FINITE TEMPERATURES\\
}

\author{\footnotesize Kensuke Inaba\footnote{Also JST, CREST, Sanbancho, Chiyoda-ku, Tokyo 102-0075, Japan}}

\address{NTT Basic Research Laboratories, NTT Corporation \\
Atsugi, Kanagawa 243-0198, Japan\\
inaba.kensuke@lab.ntt.co.jp}

\author{Sei-ichiro Suga}

\address{Department of Materials Science and Chemistry, University of Hyogo \\
 Himeji, Hyogo 671-2280, Japan\\
suga@eng.u-hyogo.ac.jp}

\maketitle

\begin{history}
\received{(Day Month Year)}
\end{history}

\begin{abstract}
We investigate the finite-temperature properties of attractive three-component (colors) fermionic atoms in optical lattices using a self-energy functional approach. 
As the strength of the attractive interaction increases in the low temperature region, a second-order transition occurs from a Fermi liquid to a color superfluid (CSF). In the strong attractive region, a first-order transition occurs from a CSF to a trionic state. In the high temperature region, a crossover between a Fermi liquid and a trionic state is observed with increasing the strength of the attractive interaction. 
The crossover region for fixed temperature is almost independent of filling. 
\end{abstract}


\section{Introduction}
Recent significant progress in cold fermionic atoms loaded in optical lattices provides us with a way to investigate the fascinating aspects of quantum many-body effects. Novel phenomena have been studied both experimentally and theoretically. A balanced mixture of three different hyperfine states of attractive $^6{\rm Li}$ fermionic atoms was succeeded in creating \cite{ottenstein,Huckans,Nakajima}. The observed three-body loss was discussed in terms of Efimov states \cite{th1,th2,th3,th4,th5}. On the other hand, theoretical studies for attractive three-component (color degrees of freedom) fermionic atoms in optical lattices were performed intensively \cite{honer1,honer2,rapp1,rapp2,Inaba}. 
It was shown that a color superfluid (CSF) emerges for the small attractive interaction, where atoms with two of three colors form the Cooper pairs and the third ones remain a Fermi liquid. As the attractive interaction becomes strong, a quantum phase transition from the CSF to the trionic state occurs \cite{rapp1,rapp2,Inaba}. In the trionic state, singlet bound states of three different color atoms are formed. Despite the high tunability of the optical lattice systems, it is difficult to cool the fermionic system because of the Pauli principle. 
It is thus desirable to investigate finite-temperature properties of the attractive three-component fermionic atoms in optical lattices.

In a recent study, we showed that a crossover occurs between the Fermi liquid and the trionic state at finite temperatures \cite{Inaba}. Furthermore, the crossover region in the phase diagram concerning temperature and interaction hardly depends on temperature at half filling. 
In real systems, a harmonic confinement potential exists, which induces a spatial change in filling. 
To observe the trionic state in experiments, the effects of temperature and filling change are indispensably investigated.

In this paper, we investigate the CSF and trionic state of attractive three-component fermionic atoms in optical lattices at finite temperatures. 
We discuss the stability of the trionic state against filling change at finite temperatures.

\section{Model and Method}
The system is assumed to keep a balanced population of fermionic atoms with three colors. The fermionic optical lattice systems can be well described by the following Hubbard-type Hamiltonian \cite{jaksch}: 
\begin{eqnarray}
{\cal H} = -t\sum_{\langle i,j \rangle, \alpha}
           c_{i\alpha}^{\dagger}c_{j\alpha}^{} 
         - \frac{U}{2}\sum_{i,\alpha \neq \beta} 
           n_{i\alpha}n_{i\beta} - \sum_{i,\alpha}\mu_{\alpha}n_{i\alpha}, 
\label{hami}
\end{eqnarray}
where $c_{i\alpha}$ annihilates a fermionic atom with color $\alpha(=1, 2, 3)$ at the $i$th lattice site, $n_{i\alpha}= c_{i\alpha}^{\dagger}c_{i\alpha}$, the subscript $\langle i,j \rangle$ indicates the sum over the nearest-neighbor sites, and $-U (<0)$ is the on-site attractive interaction between two atoms with different colors.

Using a self-energy functional approach (SFA) \cite{Potthoff03a,Potthoff03b}, we elucidate characteristics of the CSF, trionic state, and Fermi liquid, and study the phase transition and crossover between them. The SFA allows us to deal efficiently with zero- and finite-temperature properties concerning the phase transition driven by correlation effects \cite{Potthoff03a,Potthoff03b}. 
Indeed, precise results related to the Mott transition in correlated electron systems were obtained as regards thermodynamic quantities, excitation spectra, and phase diagrams. 
This method is based on the Luttinger-Ward variational method. The framework of this variational approach enables us to introduce a proper reference system which has to include the same interacting Hamiltonian as that of the original Hamiltonian. Here, we apply the two-site Anderson impurity model as the reference system. 
Parameters of the reference system are determined by the variational method. 
In the SFA \cite{Potthoff03a,Potthoff03b}, the grand potential $\Omega$ can be expressed as a function of the self-energy of the reference system $\Sigma_{\bf t}$. From the condition that the derivative of the grand potential with respect to the variational parameters becomes zero, $\partial\Omega/\partial {\bf t}=(\partial\Sigma_{\bf t}/\partial {\bf t})(\partial\Omega/\partial\Sigma_{\bf t})=0$, we obtain the reference self-energy $\Sigma_{\bf t}$ which properly describes the original correlated system, where ${\bf t}$ is the parameter matrix of the noninteracting part of the reference Hamiltonian.

We use a semicircular density of states obtained for the infinite-dimensional Bethe lattice. We calculate the CSF order parameter $\Phi=\langle c_{i1}^{\dagger}c_{i2}^{\dagger} \rangle$, the quasiparticle weight $Z=[1-\partial \Sigma_{\bf t}(\omega)/\partial \omega]|_{\omega=0}^{-1}$, which is independent of color, and the entropy per site $S/L=-\partial (\Omega/L)/\partial T$, where $L$ is the number of lattice sites. 
We further calculate the expectation value of the trion number per site $N_t=\langle N_i \rangle$ and the local susceptibility of trion-number fluctuations defined by $\chi=\int e^{i\omega \tau} \langle :N_i(\tau): :N_i(0): \rangle d\tau |_{\omega=0}$, where $:N_i(\tau):=N_i(\tau)-N_t$ with $N_{i}=n_{i1}n_{i2}n_{i3}$. Note that $\Phi$, $N_t$, and $\chi$ are independent of $i$ and color in the SFA procedure. 
We do not consider the possibility of the trionic density-wave state induced by spatial fluctuations. 
The hopping integral $t$ is used in units of energy.

\section{Results and Discussions}
\begin{figure}[th]
\centerline{\psfig{file=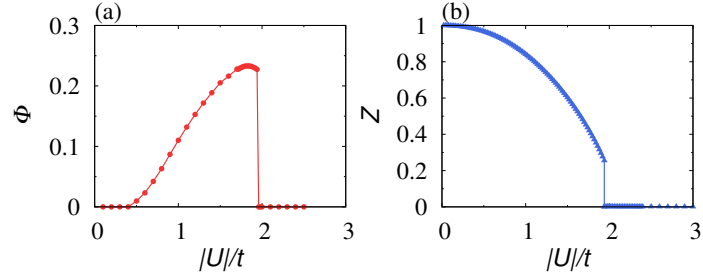,width=10cm}}
\vspace*{8pt}
\caption{$|U|/t$ dependence of (a) the CSF order parameter $\Phi$ and (b) the quasiparticle weight $Z$ at $T=0$.}
\label{f1}
\end{figure}
We begin our discussions with zero-temperature properties. The chemical potential is set at $\mu=U$ so that particle-hole symmetry can be satisfied and that a balanced population of each color atoms can be achieved. 
Figure \ref{f1} shows the CSF order parameter $\Phi$ and the quasiparticle weight $Z$ as a function of $|U|/t$. 
As $|U|/t$ is increased, $\Phi$ first increases and vanishes discontinuously at $U/t=-1.94$. 
The quasiparticle weight $Z$ decreases significantly and then drops discontinuously to zero at the same $U/t=-1.94$. The results indicate that the CSF disappears as a result of a first-order Mott transition driven by the renormalization effects. 
%
\begin{figure}[th]
\centerline{\psfig{file=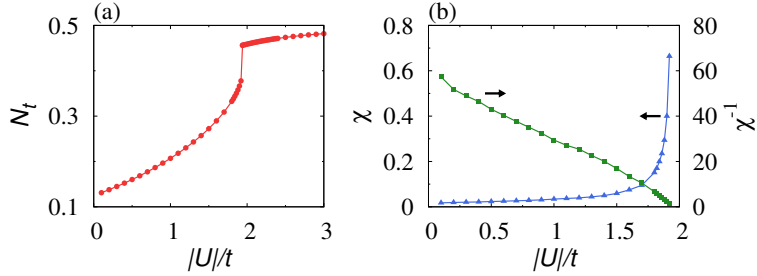,width=10cm}}
\vspace*{8pt}
\caption{$|U|/t$ dependence of (a) the expectation value of the trion number $N_t$ and (b) the local susceptibility of trion-number fluctuations $\chi$ at $T=0$.}
\label{f2}
\end{figure}
In Fig. \ref{f2}(a), we show the expectation value of the trion number $N_t$ as a function of $|U|/t$. 
At half filling, the maximum of $N_t$ is 0.5. As $|U|/t$ increases, $N_t$ increases and jumps to $\sim 0.5$ at the same critical $U/t$. 
We thus conclude that the first-order CSF-trion quantum phase transition occurs at $U/t=-1.94$. 
These results are in contrast with those in one-dimensional systems, where the CSF disappears at zero temperature because of strong quantum fluctuations and the three-particle density-wave states caused by spatial fluctuations appear \cite{capponi,molina,azaria,kantian}.

\begin{figure}[th]
\centerline{\psfig{file=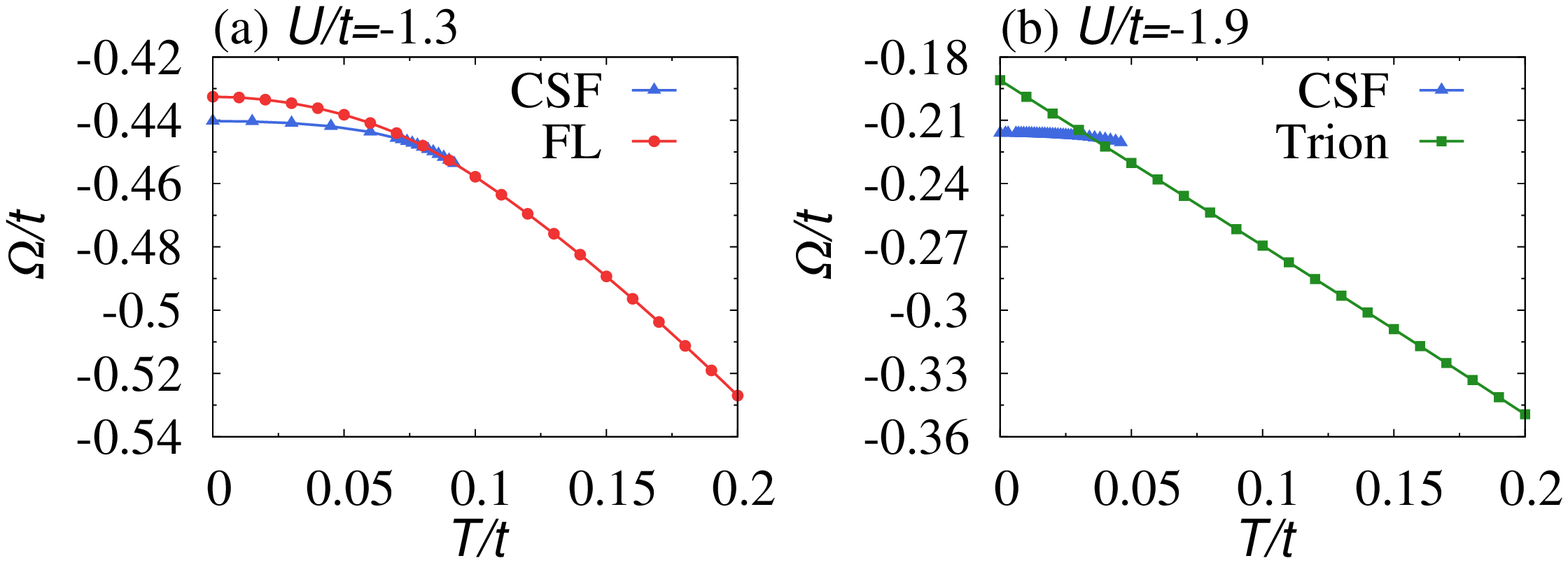,width=10cm}}
\vspace*{8pt}
\caption{Grand potentials for (a) $U/t=-1.3$ and (b) $U/t=-1.9$ as a function of temperature. FL means Fermi liquid.}
\label{f3}
\end{figure}
We investigate the phase transition and crossover at finite temperatures. 
In Fig. \ref{f3}(a) and (b), we show the grand potentials for $U/t=-1.3$ and $U/t=-1.9$ as a function of temperature. In the low temperature region for $U/t=-1.3$, the grand potential of the CSF is lower than that of the Fermi liquid. When temperature is increased, both grand potentials agree continuously. Therefore, for $U/t=-1.3$ the second-order phase transition occurs from the CSF to the Fermi liquid with increasing temperature. 
For $U/t=-1.9$, the grand potential of the CSF is lower than that of the trionic state in the low temperature region. When temperature is increased, both grand potentials cross, indicating that a first-order phase transition occurs from the CSF to the trionic state.

\begin{figure}[th]
\centerline{\psfig{file=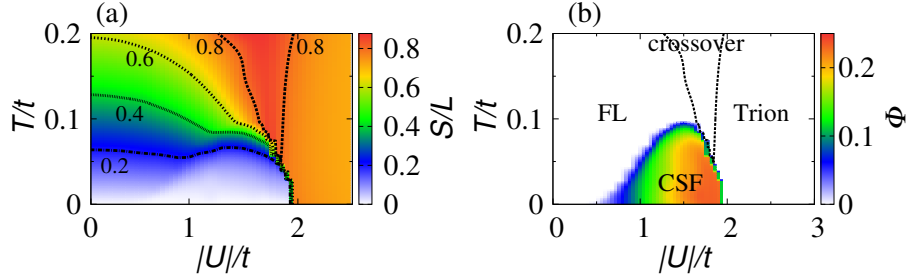,width=12cm}}
\vspace*{8pt}
\caption{(a) Contour lines of the entropy $S/L$ in the $T$-$|U|$ plane. (b) Phase diagram of the isotropic attractive three-component Hubbard model at half filling. FL means Fermi liquid.}
\label{f4}
\end{figure}
In Fig. \ref{f4}(a), the contour lines of the entropy are shown in the $T-|U|$ plane. For $|U|/t<1.7$, we find the cusp in each line. This is a manifestation of the second-order Fermi liquid-CSF transition. 
For $T/t>0.08$ around $|U|/t=1.7$, we find the large entropy region where $S/L$ is above 0.8. When $|U|/t$ is increased, the entropy decreases continuously to $S/L \sim \ln 2 \sim0.693$. Accordingly, a crossover occurs from the Fermi liquid to the trionic state. We summarize the results calculated for various parameters in the phase diagram for the Fermi liquid, CSF, and trionic state shown in Fig. \ref{f4}(b). The second-order phase transition occurs between the Fermi liquid and the CSF, and the first-order phase transition occurs between the CSF and the trionic state. There occurs a crossover between the Fermi liquid and the trionic state. 
The crossover region seems to be $|U|/t \sim 2$. This is because the crossover occurs when the trionic excitation gap ($2|U|$) \cite{Inaba} and the bandwidth ($4t$) become comparable to each other. Because of thermal fluctuations, the crossover region spreads gradually with increasing $T/t$. 
In the phase diagram, the amount of the CSF order parameter $\Phi$ is also shown. As $|U|/t$ is increased, $\Phi$ increases gradually from zero and vanishes suddenly from the value close to the maximum. This result suggests that the CSF is suppressed by the trion formation. In a recent paper, we demonstrated that, when the anisotropy of the attractive interactions are included, the CSF phase extends actually towards the large $|U|/t$ and $T/t$ region because of the suppression of the trion formation \cite{Inaba}.

We now discuss the origin of $S/L=\ln 2$ in the trionic state at half filling. In Fig. \ref{f2}(b), we show the local susceptibility of trion-number fluctuations $\chi$ at $T=0$.  When $|U|/t$ approaches $1.94$ from below, $\chi$ shows divergence. This divergence is caused by the formation of a localized trion. 
Within the SFA, the trionic state can be regarded as the Mott insulating state as shown in Fig. \ref{f1}(b), where the localized trion is formed at a site. This localized trion induces two degrees of freedom whether or not the trion is found at a site. Therefore, the entropy in the trionic state is $\ln 2$ in $T=0$ at half filling. 
A similar behavior is seen in the spin susceptibility of strongly correlated electron systems. When the two-component repulsive Hubbard model is treated by a dynamical mean-field theory, the local spin susceptibility diverges with the system approaching the Mott transition point from the weakly interacting region \cite{Georges}.  This is because a localized $S=1/2$ free spin is formed in the Mott insulating state in the infinite dimensional system, yielding the residual entropy $\ln 2$.

The residual entropy is caused by neglecting spatial fluctuations in the present SFA calculation. 
The present calculation is valid for the infinite dimensional system or for the high temperature region, where spatial fluctuations are suppressed. 
By contrast, in the finite dimensional system, spatial fluctuations cause sometimes relevant effects at low temperatures \cite{Georges}. 
As mentioned before, the trionic density-wave state is expected to appear in low temperatures at half filling. 
Furthermore, the trion hopping is expected to be induced by spatial fluctuations. The effective hopping integral and the effective nearest-neighbor repulsion can be derived as $t_{\rm eff} \sim t^3/U^2$ and $U_{\rm eff} \sim t^2/|U|$ for large $|U|/t$, respectively \cite{rapp2}. Therefore, the relation $t_{\rm eff} < U_{\rm eff}$ is satisfied, suggesting strongly correlated trions. 
It is thus expected that in the finite dimensional systems the trionic density-wave state appears close to half filling in low temperatures and the strongly renormalized trionic Fermi liquid appears in $T<t_{\rm eff}$ away from half filling. For low filling, the effective repulsion may become irrelevant and the strong renormalization effects may be suppressed. 
The investigation of the present system including the effects of spatial fluctuations is our future study.

\begin{figure}[th]
\centerline{\psfig{file=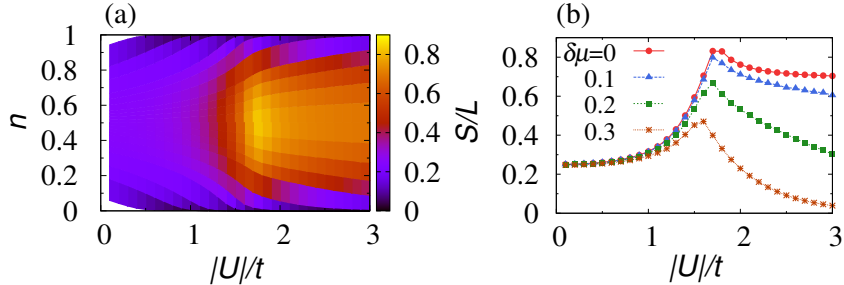,width=12cm}}
\vspace*{8pt}
\caption{Entropy $S/L$ at $T/t=0.08$ as a function of $|U|/t$ (a) for filling $0<n<1$, where $n=0.5$ denotes half filling, and (b) for fixed  $\delta \mu$.}
\label{f5}
\end{figure}
Finally, we investigate the crossover between the Fermi liquid and the trionic state for various filling with a balanced population of each color atoms. We calculate the entropy for $0<n<1$, where $n$ is filling and $n=0.5$ denotes half filling. To obtain the accurate results, we set at $T/t=0.08$ and neglect the possibility of the CSF. 
The results are shown in Fig. \ref{f5} as a function of $|U|/t$. 
In Fig. \ref{f5}(a), we find that the maximum of the entropy $S/L$ appears in $0.1<n<0.9$. When $|U|/t$ is increased in this region, $S/L$ for fixed $n$ increases gradually and takes a maximum in $1.6<|U|/t<1.7$. After the maximum, $S/L$ decreases gradually, which indicates a crossover between the Fermi liquid and the trionic state. 
The results demonstrate that the crossover region appears in $1.6<|U|/t<1.7$ irrespective of filling for fixed temperature. 
In Fig. \ref{f5}(b), we show $S/L$ for fixed $\mu$, where $\delta \mu=\mu-U$ is the shift of the chemical potential from that for half filling. 
For half filling $(\delta \mu=0)$, $S/L$ approaches $\ln 2$ with increasing $|U|/t$. For $\delta \mu \neq0$, $S/L$ decreases after a maximum. This is because particle-hole symmetry of the trion is broken and filling decreases with increasing $|U|/t$ for fixed $\delta \mu \neq0$. 
\begin{table}[pt]
\caption{$|U|/t$ and $n$ of the maximum shown in Fig. \ref{f5}(b).}
\begin{center}
{\begin{tabular}{@{}ccc@{}} \toprule
$\delta \mu$ & $|U|/t$ & $n$ \\
\colrule
0 & 1.7 & 0.5 \\
0.1 & 1.7 & 0.64\\
0.2 & 1.7 & 0.77 \\
0.3 & 1.6 & 0.86\\ \botrule
\end{tabular} }
\end{center}
\end{table}
In Tab. 1, we show $|U|/t$ and $n$ of the maximum shown in Fig. \ref{f5}(b). 
We find that $|U|/t$ of the maximum is almost independent of filling.  
In other words, the attractive interaction of the crossover for fixed temperature is almost independent of filling.

In real systems, a harmonic confinement potential exists, which yields a spatial change in filling. The center of the confinement potential is so dense. The results in Fig. \ref{f5}(a) effectively presents the effects of a confinement potential. Decrease in $n$ denotes that the observed position becomes distant from the center of the confinement potential. 
It is expected that the trionic state appears in the rather wide spatial range corresponding to $0.1<n<0.9$. 
We have shown that the crossover between the Fermi liquid and the trionic state is almost independent of filling. 
We expect that in real experiments the crossover occurs at nearly the same interaction irrespective of spatial change in filling. In addition, as shown in Fig. \ref{f4}(b), the crossover occurs at $|U|/t \sim 1.5-2.0$ irrespective of temperature. 
These findings are useful for detecting the trionic state and the crossover.

The large entropy $\sim \ln 2$ of the trionic state close to half filling suggests the possibility of the adiabatic cooling during the lattice loading caused by the Pomeranchuk effect \cite{Richardson,Werner}. However, the entropy of the trionic state decreases away from half filling. It is interesting to investigate the competition between the adiabatic cooling and the reduction of the entropy caused by the confinement potential. This issue constitutes our future study.

\section{Summary}
We have investigated the attractive three-component fermionic atoms in optical lattices at zero and finite temperatures. 
We have shown that the attractive interaction of the crossover region between the Fermi liquid and the trionic state for given temperature is almost independent of filling. 
The present results are complementary to our previous results \cite{Inaba}. We hope that our results will contribute to the study of the exotic phases of multicomponent cold fermionic atoms in optical lattices.

\section*{Acknowledgments}
Numerical computations were carried out at the Supercomputer Center, the Institute for Solid State Physics, University of Tokyo. 
This work was supported by Grants-in-Aid for Scientific Research (C) (No. 20540390) from the Japan Society for the Promotion of Science and on Innovative Areas (No. 21104514) from the Ministry of Education, Culture, Sports, Science and Technology.



\begin{thebibliography}{00}

\bibitem{ottenstein} T. B. Ottenstein, T. Lompe, M. Kohnen, A. N. Wenz, and S. Jochim, {\it Phys. Rev. Lett.} {\bf 101} (2008) 203202.

\bibitem{Huckans} J. H. Huckans, J. R. Williams, E. T. Hazlett, R. W. Stites, and K. M. O'Hara, {\it Phys. Rev. Lett.} {\bf 102} (2009) 165302. 

\bibitem{Nakajima} S. Nakajima, M. Horikoshi, T. Mukaiyama, P. Naidon, and M. Ueda, {\it Phys. Rev. Lett.} {\bf 105} (2010) 023201.

\bibitem{th1} S. Floerchinger, R. Schmidt, and C. Wetterich, {\it Phys. Rev.} A {\bf 79} (2009) 053633. 

\bibitem{th2} E. Braaten, H.-W. Hammer, D. Kang, and L. Platter, {\it Phys. Rev. Lett.} {\bf 103} (2009) 073202. 

\bibitem{th3} P. Naidon and M. Ueda, {\it Phys. Rev. Lett.} {\bf 103} (2009) 073203.

\bibitem{th4} A. N. Wenz, T. Lompe, T. B. Ottenstein, F. Serwane, G. Zurn, and S. Jochim, {\it Phys. Rev.} A {\bf 80} (2009) 040702(R).  

\bibitem{th5} E. Braaten, H.-W. Hammer, D. Kang, and L. Platter, {\it Phys. Rev.} A {\bf 81} (2010) 013605.  

\bibitem{honer1} C. Honerkamp and W. Hoftetter, {\it Phys. Rev. Lett.} {\bf 92} (2004) 170403.

\bibitem{honer2} C. Honerkamp and W. Hoftetter, {\it Phys. Rev.} B {\bf 70} (2004) 094521.

\bibitem{rapp1} A. Rapp, G. Zar\'{a}nd, C. Honerkamp, and W. Hoftetter, {\it Phys. Rev. Lett.} {\bf 98} (2007) 160405.

\bibitem{rapp2} A. Rapp, W. Hoftetter, and G. Zar\'{a}nd, {\it Phys. Rev.} B. {\bf 77} (2008) 144520.

\bibitem{Inaba} K. Inaba and S. Suga, {\it Phys. Rev.} A {\bf 80} (2009) 041602(R). 


\bibitem{jaksch} D. Jaksch, C. Bruder, J. I. Cirac, C. W. Gardiner, and P. Zoller, {\it Phys. Rev. Lett.} {\bf 81} (1998) 3108.

\bibitem{Potthoff03a} M. Potthoff, Eur. Phys. J. B. {\bf 32} (2003) 429. 

\bibitem{Potthoff03b} M. Potthoff, Eur. Phys. J. B. {\bf 36} (2003) 335. 

\bibitem{capponi} S. Capponi, G. Roux, P. Lecheminant, P. Azaria, E. Boulat, and S. R. White, {\it Phys. Rev.} A. {\bf 77} (2008) 013624.

\bibitem{molina} R. A. Molina, J. Dukelsky, and P. Schmitteckert, {\it Phys. Rev.} A {\bf 80} (2009) 013616.

\bibitem{azaria} P. Azaria, S. Capponi, and P. Lecheminant, {\it Phys. Rev.} A {\bf 80} (2009) 041604(R).

\bibitem{kantian} A. Kantian, M. Dalmonte, S. Diehl, W. Hofstetter, P. Zoller, and A. J. Daley, {\it Phys. Rev. Lett.} {\bf 103} (2009) 240401.


\bibitem{Georges} 
A. Georges, G. Kotliar, W. Krauth, and M. J. Rozenberg, {\it Rev. Mod. Phys.} {\bf 68} (1996) 13.


\bibitem{Richardson} 
R. C. Richardson, {\it Rev. Mod. Phys.} {\bf 69} (1997) 683. 

\bibitem{Werner} 
F. Werner, O. Parcollet, A. Georges, and S. R. Hassan, {\it Phys. Rev. Lett.} {\bf 95} (2005) 056401.




\end{thebibliography}
\end{document}